\documentclass[12pt]{iopart}
\usepackage{graphicx}
\begin{document}

\article[Breaking down the Fermi acceleration with inelastic
collisions]{FAST TRACK COMMUNICATIONS}{Breaking down the Fermi
acceleration with inelastic collisions}

\author{Edson D.\ Leonel}

\address{Departamento de Estat\'{i}stica, Matem\'atica Aplicada e
Computa\c c\~ao -- IGCE -- Universidade Estadual Paulista -- UNESP --
Av. 24A, 1515 -- Bela Vista -- 13.506-900 -- Rio Claro -- SP -- Brazil}

\begin{abstract}
The phenomenon of Fermi acceleration is addressed for a dissipative
bouncing ball model with external stochastic perturbation. It is shown
that the introduction of energy dissipation (inelastic collisions of
the particle with the moving wall) is a sufficient condition to break
down the process of Fermi acceleration. The phase transition from
bounded to unbounded energy growth in the limit of vanishing
dissipation is characterized.
\end{abstract}

\submitto{\JPA}



\noindent The phenomenon of Fermi acceleration (FA) is a process in
which a classical particle acquires unbounded energy from collisions
with a heavy moving wall \cite{ref1}. This phenomenon was originally
proposed by Enrico Fermi \cite{ref2} as a possible explanation for the
origin of the large energies of cosmic particles. His original model
was later modified by several investigators and applied in several
fields of physics, including plasma physics \cite{ref3}, astrophysics
\cite{ref4,ref5}, atomic physics \cite{ref6}, optics
\cite{ref7,add1,ref8} and the well known time-dependent billiard
problems \cite{ref9,ref10}. Since the seminal paper of Hammersley
\cite{ref11}, it is known that the particle's average energy grows
with time when there is a random perturbation at each impact with the
wall. This result was also confirmed for a stochastic version of the
one-dimensional bouncing ball model (a classical particle confined in
and hitting two rigid walls; one of them with a fixed position and the
other one with periodic movement) under the framework of random shift
perturbation \cite{ref12,ref13}.

One of the most important questions on Fermi acceleration is whether it
can result from the nonlinear dynamics in the absence of a random
component. The answer to this question depends on the model under
consideration. For example, for a bouncer model (a particle hitting a
periodically moving platform in the presence of a constant
gravitational field), there are specific ranges of control parameters
and initial conditions that lead to Fermi acceleration \cite{ref14}; FA
occurs when there are no invariant spanning curves on the phase space
limiting the chaotic sea and, consequently, the particle's energy gain
\cite{ref12,ref13}. For two-dimensional time-dependent billiards
(billiards with moving boundaries), the emergence of FA depends on the
type of phase space of the corresponding static version of the problem
\cite{ref15}. Therefore, as conjectured by Loskutov and collaborators
\cite{ref15}, the chaotic dynamics of a particle for static boundary
is a sufficient condition to produce FA if a time perturbation in the
boundary is introduced.

A second and very important question is: in the classical billiard
problems where FA is present, are inelastic collisions of the particle
with the boundaries a sufficient condition to suppress the unlimited
energy gain?

In this Letter, we consider the one-dimensional Fermi accelerator model
under stochastic perturbation and we seek to understand and describe a
mechanism to suppress the FA. The model consists of a classical
particle confined to bounce between two rigid and infinitely heavy
walls. One of them is fixed while the other one moves randomly with
dimensionless amplitude of motion $\epsilon$ \cite{ref16}. It is
assumed that collisions with the moving wall are inelastic, so that the
particle experiences a fractional loss of energy upon each collision. A
restitution coefficient $\alpha\in[0,1]$ controls the strength of the
dissipation. For $\alpha=1$, all collisions are elastic and therefore
FA is observed \cite{ref14}. On the other hand, for $\alpha<1$ the
model is dissipative and, as shown here, there is a bound to the energy
growth. In other words, inelastic collisions break down the FA process.
A phase transition from bounded to unbounded energy growth is observed
when the control parameter $\alpha$ approaches the unity. Here we
investigate this phase transition by studying the limit
$\alpha\rightarrow 1$. The present approach can be useful as a
mechanism to subdue the phenomenon of FA in time-dependent billiard
problems.

The dynamics of the problem is given by a two-dimensional, nonlinear
mapping for the particle velocity $V$ and the time $t$ at each impact
of the particle with the moving wall. We investigate in this Letter a
simplified version of the model \cite{ref12,ref13,ref17,ref18,ref19}
which speeds up the numerical simulations significantly without
affecting the universality class. However, similar results would
indeed be obtained for the full model. The simplified version assumes
that both walls are fixed but that, when the particle hits one of them,
it exchanges energy and momentum as if the wall were moving randomly.
Thus, considering dimensionless variables and taking into account the
inelastic collisions, the mapping that describes the dynamics of the
model is
\begin{equation}
\left\{\begin{array}{ll}
V_{n+1}=|\alpha V_n-(1+\alpha)\epsilon\sin(\phi_{n+1})|\\
\phi_{n+1}=\phi_n+{{2}\over{V_n}}+Z(n)~~{\rm mod}~~2\pi\\
\end{array}
\right.~,
\label{eq1}
\end{equation}
where $n$ is the iteration number and $Z(n)\in[0,2\pi)$ corresponds to
a random shift in the phase of the moving wall. It is clear that the
model has two control parameters, namely $\epsilon$ and $\alpha$, whose
effect must be considered.

\begin{figure}[t]
\centerline{\includegraphics[width=0.5\linewidth]{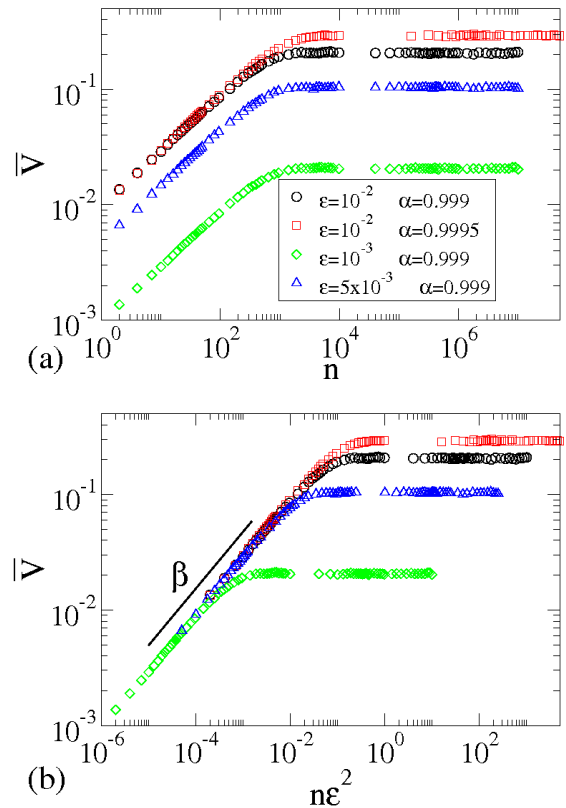}}
\caption{(Color online) (a) Behavior of ${\overline V}$ as a function
of the iteration number $n$ for different values of the control
parameters $\epsilon$ and $\alpha$. (b) Curves shown in (a) after the
change of variables $n\rightarrow n\epsilon^2$.}
\label{Fig1}
\end{figure}

The most natural observable in problems involving FA is the average
velocity, which is calculated here in two steps. The first step
consists in averaging the velocity over the orbit for a single initial
condition. It is defined as
\begin{equation}
V_i(n,\epsilon,\alpha)={{1}\over{n+1}}\sum_{j=0}^nV_{j,i}~,
\label{eq2}
\end{equation}
where the index $j$ refers to the $j$th iteration of the sample $i$.
The second step is to calculate the average over an ensemble of $M$
different initial conditions
\begin{equation}
{\overline V}={{1}\over{M}}\sum_{i=1}^MV_i~.
\label{eq3}
\end{equation}
The behavior of the average velocity ${\overline V}$ for different
values of the control parameters is shown in
Fig. \ref{Fig1}.
Note that the growth of ${\overline V}$ with $n$ is described by
different curves for different values of $\epsilon$ (Fig.
\ref{Fig1}(a)). However, a transformation $n\rightarrow n\epsilon^2$
coalesces all curves, so that they grow together for small $n$, as
seen in Fig. \ref{Fig1}(b). Also, instead of studying the behavior of
${\overline V}$ as a function of the damping coefficient $\alpha$, we
adopt the variable $(1-\alpha)$ in order to bring the transition to the
origin. Such a transformation improves visualization in log-log plots.

Let us now discuss the behavior observed in Fig. \ref{Fig1}. It is
easy to see that all curves start growing together for small iteration
numbers and then they bend towards a regime of convergency. Such a
regime is marked by a constant plateau for the average velocity. The
change from growth to saturation is characterized by a typical
crossover iteration number $n_x$. Based on the results seen in
Fig. \ref{Fig1}, one concludes that the velocity grows as a power law
of the type ${\overline V}\propto (n\epsilon^2)^{\beta}$ for 
$n\ll n_x$. For large iteration numbers, the saturation velocity for a
fixed damping coefficient $\alpha$ is given by ${\overline V}_{\rm
sat}\propto \epsilon^{\alpha_1}$. On the other hand, for a fixed
$\epsilon$, the values obtained are ${\overline V}_{\rm sat}\propto
(1-\alpha)^{\alpha_2}$. The typical crossover iteration number that
marks the transition from growth to the saturation is assumed to be of
the following form: (i) for a constant $\alpha$, $n_x\epsilon^2\propto
\epsilon^{z_1}$ and; (ii) for a fixed $\epsilon$, $n_x\epsilon^2\propto
(1-\alpha)^{z_2}$. These initial assumptions allow us to propose the
following scaling hypotheses:
\begin{itemize}
\item{For small iteration numbers ($n\ll n_x$), the average velocity
grows as
\begin{equation}
{\overline V}\propto [n\epsilon^2]^{\beta}~,
\label{eq4}
\end{equation}
where $\beta$ is a critical exponent.
}
\item{For large iteration numbers ($n\gg n_x$), the constant plateau of
the velocity is given by
\begin{equation}
{\overline V}_{\rm sat}\propto
\epsilon^{\alpha_1}(1-\alpha)^{\alpha_2}~,
\label{eq5}
\end{equation}
where both $\alpha_1$ and $\alpha_2$ are critical exponents.
}
\item{The crossover iteration number is given by
\begin{equation}
n_x\epsilon^2\propto \epsilon^{z_1}(1-\alpha)^{z_2}~,
\label{eq6}
\end{equation}
with $z_1$ and $z_2$ being the dynamical exponents.
}
\end{itemize}

\begin{figure}[b]
\centerline{\includegraphics[width=0.5\linewidth]{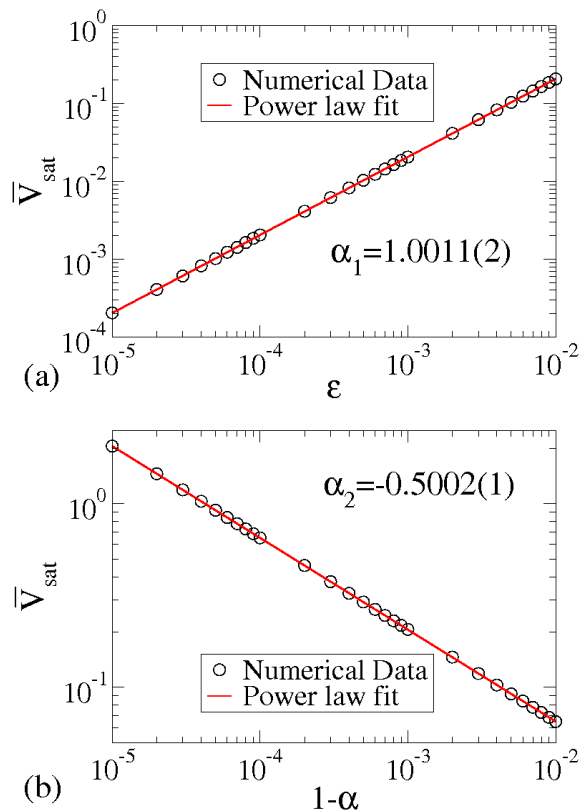}}
\caption{(Color online) (a) Plot of ${\overline V}_{\rm sat}\times
\epsilon$. A power law fitting gives $\alpha_1=1.0011(2)$. (b) Plot of
${\overline V}_{\rm sat}\times (1-\alpha)$. The exponent obtained via a
power-law fitting is $\alpha_2=-0.5002(1)$.}
\label{Fig2}
\end{figure}

These three hypotheses allow us to formally describe the average
velocity using a scaling function of the type
\begin{equation}
{\overline
V}(n\epsilon^2,\epsilon,(1-\alpha))=l{\overline V}(l^an\epsilon^2,
l^b\epsilon ,
l^c(1-\alpha))~,
\label{eq7}
\end{equation}
where $l$ is a scaling factor and $a$, $b$ and $c$ are the scaling
exponents which must be related to the critical exponents $\beta$,
$\alpha_1$ and $\alpha_2$, $z_1$ and $z_2$. Since $l$ is a scaling
factor, we can choose $l=[n\epsilon^2]^{-1/a}$. Using this expression
for $l$, Eq. (\ref{eq7}) is rewritten as
\begin{equation}
{\overline
V}(n\epsilon^2,\epsilon,(1-\alpha))=[n\epsilon^2]^{-{{1}\over { a } } }
V_1([n\epsilon^2]^{-{{b}\over{a}}}\epsilon,[n\epsilon^2]^{-{{c}\over{a}
} } (1-\alpha))~,
\label{eq8}
\end{equation}
where the function $V_1$ is assumed to be constant for $n\ll n_x$.
Comparing equations (\ref{eq8}) and (\ref{eq4}), we obtain
$\beta=-1/a$. After conducting extensive numerical simulations, it was
found that $\beta=0.494(1)\cong 0.5$, which implies $a=-2$. Let us
now consider the case $n\gg n_x$. This case allows us to choose two
distinct values for $l$, namely: (a) $l=\epsilon^{-1/b}$ and (b)
$l=(1-\alpha)^{-1/c}$. For case (a), the scaling function is rewritten
as
\begin{equation}
{\overline
V}(n\epsilon^2,\epsilon,(1-\alpha))=\epsilon^{-{{1}\over{b}}}
V_2(\epsilon^{-{{a}\over{b}}}n\epsilon^2,\epsilon^{-{{c}\over{b}}}
(1-\alpha))~,
\label{eq9}
\end{equation}
with the function $V_2$ being constant for $n\gg n_x$ and $\alpha$
constant. A comparison of equations (\ref{eq9}) and (\ref{eq5})
furnishes $\alpha_1=-1/b$. After fitting a power law to data on the
plot ${\overline V}_{\rm sat}\times \epsilon$, we find
$\alpha_1=1.0011(2)\cong 1$ (see Fig. \ref{Fig2}(a)), which provides
$b=-1$.
We now have to consider case (b), i.e., $l=(1-\alpha)^{-1/c}$. Using
this scaling factor, Eq. (\ref{eq7}) is given by
\begin{equation}
{\overline
V}(n\epsilon^2,\epsilon,(1-\alpha))=(1-\alpha)^{-{{1}\over{c}} }
V_3((1-\alpha)^{-{{a}\over{c}}}n\epsilon^2,(1-\alpha)^{-{{b}\over{c}}}
\epsilon)
\label{eq10}
\end{equation}
where we assume that $V_3$ is constant for $n\gg n_x$ and $\epsilon$
constant. Comparing equations (\ref{eq10}) and (\ref{eq5}), it is easy
to see that $\alpha_2=-1/c$. A power-law fitting to the data for
${\overline V}_{\rm sat}\times (1-\alpha)$ gives that
$\alpha_2=-0.5002(1)\cong -0.5$, yielding $c=2$.

\begin{figure}[b]
\centerline{\includegraphics[width=0.5\linewidth]{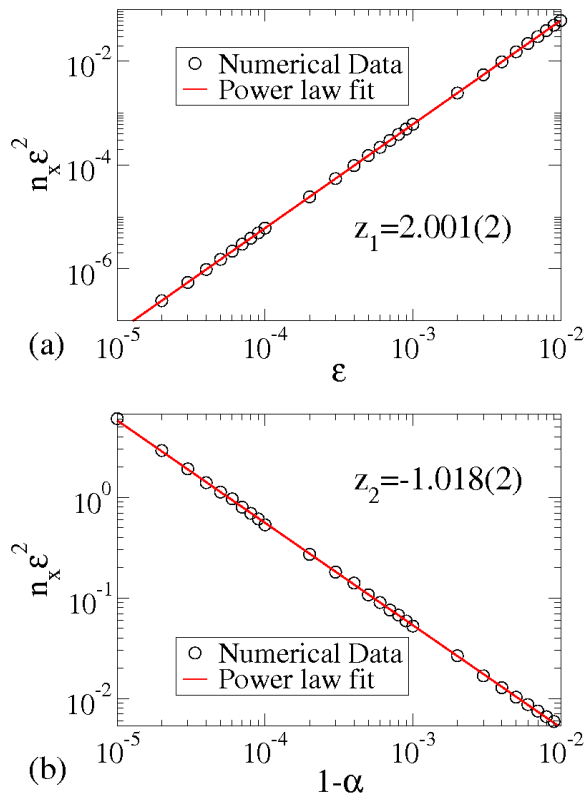}}
\caption{(Color online) (a) Plot of $n_x\epsilon^2\times\epsilon$.
After fitting a power law we obtain that $z_1=2.001(2)$. (b) Plot of
$n_x\epsilon^2\times (1-\alpha)$ for a fixed $\epsilon=10^{-2}$.
Fitting a power law, we find $z_2=-1.018(2)$.}
\label{Fig3}
\end{figure}

Considering the different expressions for the scaling factor $l$
obtained for $n\ll n_x$ and case (a) of $n\gg n_x$, we obtain
\begin{equation}
n\epsilon^2=\epsilon^{{a}\over{b}}~.
\label{eq11}
\end{equation}
A comparison of equations (\ref{eq11}) and (\ref{eq6}) provides
$z_1=a/b$. Fitting a power law to the data on the plot
$n_x\epsilon^2\times \epsilon$ furnishes $z_1=2.001(2)$, as shown in
Fig. \ref{Fig3}(a).
This result is in agreement with the ratio $a/b=2$. As a next step, we
consider the case of the different expressions for the scaling factor
$l$ obtained for $n\ll n_x$ and case (b) of $n\gg n_x$. Such procedure
gives
\begin{equation}
n\epsilon^2=(1-\alpha)^{{a}\over{c}}~.
\label{eq12}
\end{equation}
Comparing equations (\ref{eq12}) and (\ref{eq6}), we obtain that
$z_2=a/c$. A power-law fitting to the data on the plot
$n_x\epsilon^2\times (1-\alpha)$ (Fig. \ref{Fig3}(b)), with
$\epsilon=10^{-2}$, gives $z_2=-1.018(2)$. This result is also in good
agreement with the result obtained from $a/c=-1$.

\begin{figure}[t]
\centerline{\includegraphics[width=0.5\linewidth]{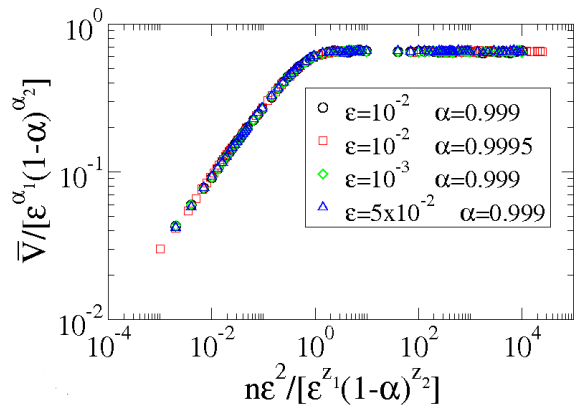}}
\caption{(Color online) Collapse of the curves shown in Fig.
\ref{Fig1}(b) onto a single and universal plot.}
\label{Fig4}
\end{figure}

Let us now discuss the consequences of the critical exponents
$\alpha_2$ and $z_2$ on the FA. It is clear that, in the limit
$\alpha\rightarrow 1$, both Eqs. (\ref{eq5}) and (\ref{eq6}) diverge.
Therefore, in the limit of vanishing dissipation, the average velocity
${\overline V}$ and the crossover iteration number $n_x$ diverge. This
is a clear signature of a phase transition from bounded to unbounded
energy growth. Finally, in order to check the validity of the scaling
hypotheses, we can now proceed to collapse all the curves onto a single
and universal plot, as shown in Fig. \ref{Fig4}.

In summary, the problem of a stochastic bouncing ball model with
inelastic collisions is addressed. It is shown that, when dissipation
is present, the average velocity of a classical particle grows with
time and then reaches a regime of saturation. However, in the limit of
vanishing dissipation, the phenomenon of Fermi acceleration is
recovered. Similar results were also observed for a dissipative and
deterministic bouncer model \cite{ref20,ref21}. These results allow us
to propose the following conjecture: {\it For one-dimensional billiard
problems that show unlimited energy growth for both their deterministic
and stochastic dynamics, the introduction of inelastic collision in the
boundaries is a sufficient condition to break down the phenomenon of
Fermi acceleration.} Such phenomenon is also expected to be observed in
two-dimensional, time-varying billiard problems since the FA mechanism
is the same as that of the one-dimensional case.

E.D.L. thanks Prof. P.V.E. McClintock for fruitful discussions and Dr.
G.J.M. Garcia for a careful review of the manuscript. Support from
CNPq, FAPESP and FUNDUNESP, Brazilian agencies, is gratefully
acknowledged.


\section*{References}

\begin{thebibliography}{9}

\bibitem{ref1} Ulam S 1961 {\it Proceedings of the Fourth Berkeley
Symposium on Math. Statistics and Probability} {\bf Vol 1} 315
(University of California Press, Berkeley)

\bibitem{ref2} Fermi E 1949 {\it Phys. Rev} {\bf 75} 1169

\bibitem{ref3} Milovanov A V, Zelenyi L M 2001 {\it Phys. Rev. E} {\bf
64} 052101

\bibitem{ref4} Veltri A, Carbone V 2004 {\it Phys. Rev. Lett.} {\bf
92} 143901

\bibitem{ref5} Kobayakawa K, Honda Y S, Samura T 2002 {\it Phys. Rev.
D} {\bf 66} 083004

\bibitem{ref6} Lanzano G {\it et al.} 1999 {\it Phys. Rev. Lett.} {\bf
83} 4518

\bibitem{ref7} Saif F, Bialynicki-Birula I, Fortunato M, Schleich W P
1998 {\it Phys. Rev. A} {\bf 58} 4779

\bibitem{add1} Saif F, Rehman I 2007 {\it Phys. Rev. A} {\bf 75} 043610

\bibitem{ref8} Steane A, Szriftgiser P, Desbiolles P, Dalibard J 1995
{\it Phys. Rev. Lett.} {\bf 74} 4972

\bibitem{ref9} Loskutov A, Ryabov A B 2002 {\it J. Stat. Phys.} {\bf
108} 995

\bibitem{ref10} Egydio de Carvalho R, de Sousa F C, Leonel E D 2006
{\it J. Phys. A} {\bf 39} 3561

\bibitem{ref11} Hammersley J M 1961 {\it Proceedings of the Fourth
Berkeley Symposium on Math. Statistics and Probability} {\bf Vol 1} 79
(University of California Press, Berkeley)

\bibitem{ref12} Karlis A K, Papachristou P K, Diakonos F K,
Constantoudis V, Schmecher P 2006 {\it Phys. Rev. Lett.} {\bf 97}
194102

\bibitem{ref13} Karlis A K, Papachristou P K, Diakonos F K,
Constantoudis V, Schmecher P 2007 {\it Phys. Rev. E} {\bf 76} 016214

\bibitem{ref14} Lichtenberg A J, Lieberman M A, Cohen R H 1980 {\it
Physica D} {\bf 1} 291

\bibitem{ref15} Loskutov A, Ryabov A B, Akinshin L G 2000 {\it J. Phys.
A} {\bf 33} 7973

\bibitem{ref16} This is an artificial mechanism commonly used to
produce Fermi acceleration in billiard problems. See for example Refs.
\cite{ref12,ref13}.

\bibitem{ref17} Lichtenberg A J, Lieberman M A 1992 {\it Regular and
Chaotic Dynamics} (Appl. Math. Sci. {\bf 38} Springer Verlag, New York)

\bibitem{ref18} Leonel E D, McClintock P V E 2005 {\it J. Phys. A} {\bf
38} 823

\bibitem{ref19} Ladeira D G, da Silva J K L 2006 {\it Phys. Rev. E}
{\bf 73} 026201

\bibitem{ref20} Leonel E D, Livorati A L P 2007 {\bf Unpublished}

\bibitem{ref21} Ladeira D G, Leonel E D 2007 {\bf Unpublished}

\end{thebibliography}
\end{document}